\documentclass[prl,aps,twocolumn,preprintnumbers,showpacs,showkeys,superscriptaddress]{revtex4}

\usepackage{epsfig,amssymb,amsfonts,verbatim}

\def\bfl{\begin{flushleft}}
\def\efl{\end{flushleft}}
\def\bfr{\begin{flushright}}
\def\efr{\end{flushright}}
\def\bc{\begin{center}}
\def\ec{\end{center}}
\def\be{\begin{equation}}
\def\ee{\end{equation}}
\def\ba{\begin{eqnarray}}
\def\ea{\end{eqnarray}}
\def\baa#1{\begin{array}{#1}}
\def\eaa{\end{array}}
\def\bw{\begin{widetext}}
\def\ew{\end{widetext}}
\def\nn{\nonumber }
\def\lb#1{\label{#1}}

\begin{document}

\title{Magneto-electronic transport theory in ferromagnets above the Curie temperature and in semiconductors}

\author{Andrew Das Arulsamy}
\affiliation{Condensed Matter Group, Division of Exotic Matter,
No. 22, Jalan Melur 14, Taman Melur, 68000 Ampang, Selangor DE,
Malaysia}

\date{\today}

\begin{abstract}
Quantitative differences of Lagrange multipliers between standard
Fermi-Dirac statistics (FDS) and Ionization energy ($E_I$) based
FDS (iFDS) are analyzed in detail to obtain reasonably accurate
interpretations without violating the standard FDS. The
resistivity and Hall-resistance models in 1D, 2D and 3D are also
derived to illustrate the transport phenomena in semiconducting
manganites. It is shown via calculation that the charge carriers
in these materials seem to be strongly correlated in term of
electron-ion attraction or simply, fermions in those materials are
somewhat gapped due to Coulomb attraction. This Coulomb attraction
naturally captures the polaronic effect in manganites. $E_I$ is
found to be the only essential parameter that predicts
$\rho(T,doping,pressure,magnetic~field)$ quite accurately.
However, this model as will be pointed out, is not suitable for
metals with free-electrons and strong electron-phonon scattering.
It is to be noted that iFDS and
$\rho(T,doping,pressure,magnetic~field)$ models are only valid in
the paramagnetic region of ferromagnetic manganites and other
doped-semiconductors. Recent X-ray photoemission spectroscopy
(XPS) studies have indicated that there exists a critical
crossover from Mn$^{3+}$ to Mn$^{2+}$ depending upon Mn's
concentrations in (Ga$_{1-x}$Mn$_x$)As diluted magnetic
semiconductors (DMS). Such phenomenon occurring at certain
critical concentration directly point towards the applicability of
$E_I$ based Fermi-liquid theory (iFLT). As such, iFDS is also
discussed with respect to DMS above the Curie temperature.
\end{abstract}

 \maketitle


\section{1. Introduction}\lb{s-in}

Doped-compounds including oxides and its electrical and magnetic
measurements have contributed enormously to the understanding of
electrical properties of ferromagnets at $T$ $>$ $T_C$
(paramagnetic $\leftrightarrow$ ferromagnetic transition $T$) and
semiconductors. The complete mechanism above $T_C$ for
ferromagnets is somewhat vague since the variation of $\rho(T,
doping, pressure, magnetic~field~({\bf H}))$ in term of hopping
activation energy, $E_p$ is still unclear. I.e., the variation of
$E_p$ with doping is not explicitly predictable. Hence, it is
essential to study and understand the variation of
$\rho(T,doping,pressure,{\bf H})$ in order to enhance the
predictability of electrical properties that may accelerate the
possible applications of these materials. Two-dimensional (2D)
$E_I$ based Fermi liquid model was originally used to describe
$c$-axis and $ab$-planes conduction of overdoped cuprate
superconductors~\cite{arulsamy1,arulsamy2,arulsamy2a,andrew2}.
Subsequently, it was further developed to capture both
$T_{crossover}$ ($c$-axis pseudogap) and $T^*$ (spin gap
characteristic $T$) in $ab$-planes peculiar conduction involving
spinons and holons, which is known as the hybrid
model~\cite{arulsamy3,arulsamy4}. Take note that there are three
typographical errors in Ref.~\cite{arulsamy3}, two of them can be
found in Eq.~(9) of Ref.~\cite{arulsamy3} where
$(m^*_em^*_h)^{-3/4}$ and $(2\pi \hbar^2/k_B)^{-3/2}$ should be
replaced with $(m^*_em^*_h)^{-1/4}$ and $(2\pi \hbar^2/k_B)^{3/2}$
respectively. Therefore, the fitting parameter, $A_2$ is actually
equals to $(A_2/2e^2)(m^*_em^*_h)^{-1/4}(2\pi \hbar^2/k_B)^{3/2}$.
One can safely claim that manganites are one of the well studied
compound after Cuprates due to its vast applications with hardly
any temperature constraints. Literally, La-Nd based manganites
have been investigated in almost all aspects of experimental
techniques, yet there are still several theoretical rendezvous on
magneto-electronic properties both below and above $T_C$. A wide
variety of these properties based on doping and Mn's valence state
in manganites were reported to understand the transport
mechanism(s)~\cite{fresard,mancini,nagaevi,horyn,aguero,chan,kumar,machado}.
Among them, the influence of grain boundary as a
barrier~\cite{ju}, as a region of depleted $T_C$~\cite{hernandez}
and polaronic effect~\cite{nagaevii,yang} on electrical properties
were reported. Direct proportionality of {\bf H} with
$T_C$~\cite{nagaeviii,nagaeviv,abramovich} and resistivity with
defects or substrate-film lattice
incompatibility~\cite{rubi,hur,heffner,prokhorov,boujelbeni,boujelbenii,sun,mei,coldea}
are also regarded as equally important to determine the electrical
properties of manganites. Furthermore, metallic conduction below
$T_C$ has been studied using double exchange mechanism (DEM)
between $s$ and $d$ orbitals~\cite{demin} and the displacement of
hysteresis loop in field-cooled sample with an additional scenario
of non-linear spin and charge fluctuations due to
magnon~\cite{solontsov}. Explanations in term of hopping electrons
and DEM~\cite{yu}, and the influence of microstructural transition
arises from ionic radius or valence state of Nd in
Nd$_x$Sm$_{1-x}$Ca$_{0.8}$MnO$_3$~\cite{filippov} were also
reported extensively.

Presently, the discovery of ferromagnetism (FM) in diluted
magnetic-semiconductors (DMS) such as Mn doped GaAs is expected to
have an enormous impact on scientific and technological
developments since both electronic and magnetic properties are
realized simultaneously. Two primary advantages of DMS over
manganites are the ease of controlling electrons' ($n$) and holes'
($p$) concentrations and electromigration hardly plays any
significant role due to zilch oxygen content. Electromigration in
oxides is a process of migration and diffusion of mainly oxygen
atoms (weakly bound) in the vicinity of high temperature, $T$
(random direction), electric, {\bf E} and magnetic fields, {\bf H}
(directional) as well as in high vacuum. The attention for the
discovery of DMS by Munekata {\it et al.}~\cite{munekata1} and
later by Ohno {\it et al.}~\cite{ohno2,ohno3} in the early 90s was
actually due to the possibility of having a brand new type of
magneto-electronic materials other than manganites both in term of
experimental facts and theories though it does not mimic the
excitement for the discovery of cuprate superconductors in the
late 80s. Nevertheless, DMS is indeed unique in a sense that it
exposes the potential applications in a wide variety of electronic
industries namely, high-density non-volatile magnetic memory
integrated with semiconductor integrated-circuits, magnetic
sensors, optical isolators integrated with semiconductor lasers
for optical communication systems~\cite{henini4} and
nanotechnology~\cite{takahashi,aguekian}. The intensive progress
of DMS's sample growth is apparently fuelled by the advancement of
Molecular-Beam-Epitaxy's (MBE) technique in the
90s~\cite{ohno,park,hayashi,yasuda}. Basically, FM arising from
the said doped material was argued against the well known Zener
DEM~\cite{zener5,zener6} since DMS requires long-range spin
interaction. This long-range spin interaction is necessary to
produce FM due to very low concentration of magnetic ion (Mn) in
DMS and consequently it is named as such. The highest Curie
temperature ($T_C$) has been reported to be at 110 K for MBE
grown, Ga$_{0.947}$Mn$_{0.053}$As compound~\cite{ohno3}. Recently
however, Teraguchi {\it et al}.~\cite{teraguchi} and Saito {\it et
al}.~\cite{saito} have highlighted that Ga$_{0.94}$Gd$_{0.06}$N
and Zn$_{0.8}$Cr$_{0.2}$Te have $T_C$s of about 300 K and 400 K
respectively. As a matter of fact, it is well known that only
oxides namely, Pr,La,Nd-Ca,Sr-Mn,Fe-O compounds provide the
established phenomenon called FM with $T_C$ ranging from above and
far below room temperature ($T_r$). In fact, the discovery of FM
in manganites were first reported by Jonker and van
Santen~\cite{jonker7,santen8} in early 50s and antiferromagnetism
(AFM) in Cd$_x$Mn$_{3-x}$O$_4$ by Dey~\cite{dey}. This particular
oxide is also captured in detail by Zener DEM~\cite{zener5,zener6}
with slight improvements in term of polaronic
effect~\cite{louca9,millis10} due to Jahn-Teller splitting of
Mn$^{3+}$ ions. However, FM with a $T_C$ up to 110 K that was
observed in Ga-Mn-As magnetic semiconductors~\cite{ohno3} have
invoked theoretical and experimental endeavors due to long-range
spin interaction.

Several competing models have attempted to describe FM in Mn doped
GaAs that includes Ruderman-Kittel-Kasuya-Yosida (RKKY) model, DEM
and antiferromagnetic (AFM) superexchange mechanism as well as the
Mn-holes (Mn-$h$) complex. Actually the RKKY approach in the
presence of the Friedel oscillations can be approximated to DEM
for DMS~\cite{dietl11}. The itinerant character of magnetic
electrons and quantum oscillations (Friedel) of the electron's
spin polarization around the localized spins were established for
the theory of magnetic metals. Hence, the resulting competition
between FM and AFM interactions give rise to spin-glass
freezing~\cite{dietl11,dietl12}. In DMS however, the mean distance
between the carriers are larger than the distance between spins.
As a consequence, the exchange interaction mediated by the
carriers in DMS favors FM rather than spin-glass freezing, which
is for magnetic metals~\cite{dietl11,dietl12}. Originally, Van
Esch {\it et al.}~\cite{esch13,esch14} have given an
interpretation for FM in DMS in term of interaction between Mn
ions and $h$ surrounding it in which, the Mn ions are neutral
Mn$^{3+}$ acceptors associated with 3$d^{5}$ + $h$ configuration.
In addition, this neutral Mn$^{3+}$ acceptors that contribute to
magnetic properties could be compensated by As, where for a higher
concentrations of Mn, instead of replacing Ga it will form a
six-fold coordinated centers with As
(Mn$^{6As}$)~\cite{esch13,esch14}. These centers will eventually
reduce the magnitude of FM in DMS due to the loss of spin-spin
interaction between Mn(3$d^{5})$ and $h$. Interestingly, Dietl
{\it et al.}~\cite{dietl11,dietl12} have successfully incorporated
both DEM and AFM superexchange mechanism into the Mn-$h$ complex
in order to capture the overall mechanism involved in III-V DMS.
Such incorporations are important since DEM seems to be the only
mechanism that predicts $T_C$ accurately after taking band
structure effect into account~\cite{dietl11,dietl12}. DMS was
assumed to be charge-transfer insulators by Dietl {\it et al.}
because $h$ from Mn$^{2+}$ does not reside on $d$ orbitals but
occupies an effective mass Bohr orbit with large intrasite
correlation energy. Moreover, Mn ions also function as a source of
localized spins as well as acceptors~\cite{dietl11}. Dietl {\it et
al}.~\cite{dietl11,dietl12} have further suggested that FM between
localized spins ($S$ = 5/2) of Mn is mediated by delocalized or
weakly localized holes surrounding it and the resulting spin-spin
interaction are indeed long range as it should be. On the other
hand, the short-range AFM superexchange (if any) arises from the
spin polarization of occupied electron bands (valence band). As
such, AFM superexchange is mediated by short-ranged localized
spins unlike FM where it requires at least weakly localized spins
for long range interaction. The readers are referred to Dietl {\it
et al}.~\cite{dietl11} for a thorough review on DMS.

Apart from that, Okabayashi {\it et al}.~\cite{okabayashi21} have
pointed out convincingly that there exists an indeterminacy on pin
pointing whether Mn$^{2+}$ (negative) or Mn$^{3+}$ (neutral) ion
that dominates Ga$_{0.93}$Mn$_{0.07}$As DMS. Both ions (Mn$^{2+}$
and Mn$^{3+}$) were found to be favorable towards FM. In order to
explain this scenario, Ando {\it et al}.~\cite{ando22} have
proposed that Mn$^{3+}$ ions populate at very low Mn
concentrations due to low screening effect or large long-range
Coulomb interaction. I.e. long-range Coulomb interaction will be
screened-out for higher Mn concentrations. This interesting effect
of transition between Mn$^{2+}$ (negative) $\leftrightarrow$
Mn$^{3+}$ (neutral) with Mn doping can also be analyzed in
accordance with iFLT. The mechanism that governs DMS above $T_C$
and the predictability of its $\rho(T$,doping) and $R_H(T$,doping)
as well as the changes in $T_C$ with {\bf H} are also evaluated.
Actually, the transition of Mn$^{2+}$ (negative) $\leftrightarrow$
Mn$^{3+}$ (neutral) can be described by simply calculating $E_I$
for each doping from its respective $\rho(T$,doping).

In this paper, iFDS is re-derived to extract the Lagrange
multipliers so as to trigger sufficient interest for applications
in other compounds such as ferromagnetic manganites and DMS above
$T_C$. $\rho(T)$ curves are simulated at various doping or gap
($E_I$), pressure and $T$ to further enhance its applicability.
Basically, we will stress the polaronic effect in semiconducting
manganites and DMS in order to obtain accurate predictions of the
charge carriers transport mechanisms above $T_C$ where the changes
in $E_I$ is well accounted for with doping. In addition, it is
also shown with detailed derivation of somewhat different Lagrange
multipliers that separately influence the standard FDS and iFDS.
In addition, the probability functions of electrons and holes,
charge carriers' concentrations, $\rho(T$,doping) and
$R_H(T$,doping) models in 1D, 2D and 3D are derived as well via
the standard quantum statistical method. Interpretations of
electrical properties based on these models for manganites and DMS
are highlighted in detail. Quantum statistical method is used
instead of other approaches because the former method will lead to
the final results to understand $\rho(T,doping,pressure,{\bf H})$
without unnecessary approximations.

\section{2. Theoretical details}

\subsection{2.1. Fermi-Dirac distribution function based on ionization energy}\lb{s-eqs}

Both FDS and iFDS are for the half-integral spin particles such as
electrons and holes. Its total wave function, $\Psi$ has to be
antisymmetric in order to satisfy quantum-mechanical symmetry
requirement. Under such condition, interchange of any 2 particles
($A$ and $B$) of different states, $\psi_i$ and $\psi_j$ ($j$
$\ne$ $i$) will result in change of sign, hence the wave function
for Fermions is in the form of

\begin {eqnarray}
\Psi_{i,j}(C_A,C_B) = \psi_i(C_A)\psi_j(C_B) - \psi_i(C_B)\psi_j(C_A),\label{eq:1}
\end {eqnarray}

The negative sign in Eq.~(\ref{eq:1}) that fulfils antisymmetric
requirement is actually due to one of the eigenvalue of exchange
operator~\cite{griffiths6}, {\bf P} = $-$1. The other eigenvalue,
{\bf P} = $+$1 is for Bosons. $C_A$ and $C_B$ denote all the
necessary cartesian coordinates of the particles $A$ and $B$
respectively. Equation~(\ref{eq:1}) is nothing but Pauli's
exclusion principle. The one-particle energies $E_1$, $E_2$,
$E_3$, ..., $E_m$ for the corresponding one-particle quantum
states $q_1$, $q_2$, $q_3$, ..., $q_m$ can be rewritten as
($E_{initial~state}$ $\pm$ $E_I)_1$, ($E_{initial~state}$ $\pm$
$E_I)_2$, ($E_{initial~state}$ $\pm$ $E_I)_3$, ...,
($E_{initial~state}$ $\pm$ $E_I)_m$. Note here that
$E_{initial~state}$ + $E_I$ = $E_{electrons}$ and
$E_{initial~state}$ $-$ $E_I$ = $E_{holes}$. Subsequently, the
latter ($E_{initial~state}$ $\pm$ $E_I)_i$ version where $i$ = 1,
2, 3, ..., $m$ with $E_I$ as an additional inclusion will be used
to derive iFDS and its Lagrange multipliers. This $\pm E_I$ is
inserted carefully to justify that an electron to occupy a higher
state $N$ from initial state $M$ is more probable than from
initial state $L$ if condition $E_I(M)$ $<$ $E_I(L)$ at certain
$T$ is satisfied. As for a hole to occupy a lower state $M$ from
initial state $N$ is more probable than to occupy state $L$ if the
same condition above is satisfied. $E_{initial~state}$ is the
energy of a particle in a given system at a certain initial state
and ranges from $+\infty$ to 0 for electrons and 0 to $-\infty$
for holes. In contrast, standard FDS only requires $E_i$ ($i$ = 1,
2, 3, ..., $m$) as the energy of a particle at a certain state.
Denoting $n$ as the total number of particles with $n_1$ particles
with energy ($E_{initial~state}$ $\pm$ $E_I)_1$, $n_2$ particles
with energy ($E_{initial~state}$ $\pm$ $E_I)_2$ and so on implies
that $n$ = $n_1$ + $n_2$ + $n_3$ + ... + $n_m$. As a consequence,
the number of ways for $q_1$ quantum states to be arranged among
$n_1$ particles is given as

\begin {eqnarray}
P(n_1,q_1) = \frac{q_1!}{n_1!(q_1 - n_1)!}.\label{eq:2}
\end {eqnarray}

Now it is easy to enumerate the total number of ways for $q$
quantum states ($q$ = $q_1$ + $q_2$ + $q_3$ + ... + $q_m$) to be
arranged among $n$ particles, which is

\begin {eqnarray}
P(n,q) = \prod\limits_{i=1}^{\infty} \frac{q_i!}{n_i!(q_i - n_i)!} .\label{eq:3}
\end {eqnarray}

The most probable configuration at certain $T$ can be obtained by
maximising $P(n,q)$ subject to the restrictive conditions

\begin {eqnarray}
&&\sum_i^{\infty} n_i = n, \nn \\&&
\sum_i^{\infty} dn_i = 0.\label{eq:4}
\end {eqnarray}

\begin {eqnarray}
&&\sum_i^{\infty} (E_{initial~state}\pm E_I)_i n_i = E, \nn \\&&
\sum_i^{\infty} (E_{initial~state}\pm E_I)_i dn_i = 0.\label{eq:5}
\end {eqnarray}

The method of Lagrange multipliers~\cite{griffiths6} can be
employed to maximise Eq.~(\ref{eq:3}). Hence, a new function,
$F(x_1, x_2, x_3,...\mu, \lambda,...)$ = $f + \mu f_1 + \lambda
f_2$ +... is introduced and all its derivatives are set to zero

\begin {eqnarray}
\frac{\partial F}{\partial x_n} = 0;~~~ \frac{\partial F}{\partial
\mu} = 0;~~~ \frac{\partial F}{\partial \lambda} = 0.\label{eq:6}
\end {eqnarray}

As such, one can let the new function in the form of

\begin {eqnarray}
F = \ln P + \mu \sum_i^{\infty} dn_i + \lambda \sum_i^{\infty}
(E_{initial~state}\pm E_I)_i dn_i.\label{eq:7}
\end {eqnarray}

After applying Stirling's approximation, $\partial F$/$\partial n_i$ can be written as

\begin {eqnarray}
\frac {\partial F}{\partial n_i} &&= \ln (q_i - n_i) - \ln n_i +
\mu + \lambda (E_{initial~state}\pm E_I)_i \nn \\&& =
0.\label{eq:8}
\end {eqnarray}

Thus, the Fermi-Dirac statistics based on ionization energy is simply given by

\begin {eqnarray}
\frac {n_i}{q_i} = \frac{1}{\exp [\mu + \lambda
(E_{initial~state}\pm E_I)_i] + 1}.\label{eq:9}
\end {eqnarray}

The importance of $E_I$'s inclusion is that it can be interpreted
as a charge gap that will be described later and also,
particularly the $E_I$ can be used to estimate the resistivity
transition upon substitution of different valence state ions.

\subsection{2.2. Lagrange multipliers}\lb{s-eqs}

By utilizing Eq.~(\ref{eq:9}) and taking $\exp[\mu + \lambda(E \pm E_I)]$ $\gg$ 1,
one can arrive at the probability function for electrons
in an explicit form as~\cite{arulsamy2}

\begin{eqnarray}
f_e({\bf k}) = \exp \left[-\mu-\lambda\left(\frac{\hbar^2{\bf k}^2}{2m}+E_I\right) \right],
\label{eq:10}
\end{eqnarray}

Similarly, the probability function for $h$ is given by

\begin{eqnarray}
f_h({\bf k}) = \exp\left[\mu + \lambda\left(\frac{\hbar^2{\bf k}^2}{2m}-E_I\right) \right].
\label{eq:11}
\end{eqnarray}

The parameters $\mu$ and $\lambda$ are the Lagrange multipliers. $\hbar$ $=$ $h/2\pi$, $h$ $=$
Planck constant and $m$ is the charge carriers' mass. Note that $E$ has been substituted
with $\hbar^2{\bf k}^2/2m$. In the standard FDS,
Eqs.~(\ref{eq:10}) and~(\ref{eq:11}) are simply given by, $f_e({\bf k})$ $=$
$\exp[-\mu-\lambda(\hbar^2{\bf k}^2/2m)]$ and $f_h({\bf k})$ $=$
$\exp[\mu+\lambda(\hbar^2{\bf k}^2/2m)]$. Equation~(\ref{eq:4}) can be rewritten
by employing the 3D density of states' (DOS) derivative, $dn$ $=$ $V{\bf k}^2d{\bf k}/2\pi^2$,
that eventually gives~\cite{griffiths6}

\begin{eqnarray}
n & = & \frac{V} {2\pi^2} e^{-\mu}\int\limits_{0}^{\infty} {\bf k}^2 \exp\left[\frac{-\lambda
\hbar^2{\bf k}^2}{2m} \right]d{\bf k},\label{eq:12}
\end{eqnarray}

\begin{eqnarray}
p = \frac{V}{2\pi^2}e^{\mu}\int\limits_{-\infty}^0 {\bf k}^2
\exp\left[\frac{-\lambda\hbar^2{\bf k}^2}{2m}\right]d{\bf k}.\label{eq:13}
\end{eqnarray}

$n$ is the concentration of electrons whereas $p$ represents holes' concentration.
$V$ denotes volume in
{\bf k}-space. The respective solutions of Eqs.~(\ref{eq:12}) and~(\ref{eq:13}) are given below

\begin{eqnarray}
\mu_e & = & -\ln\left[\frac{n}{V}\left(\frac{2\pi\lambda\hbar^2}{m}\right)^{3/2}\right],
\label{eq:14}
\end{eqnarray}

\begin{eqnarray}
\mu_h & = & \ln\left[\frac{p}{V}\left(\frac{2\pi\lambda\hbar^2}{m}\right)^{3/2}\right].
\label{eq:15}
\end{eqnarray}

The subscripts $e$ and $h$ represent electrons and holes respectively. Separately,
Eq.~(\ref{eq:5}) after neglecting $\pm E_I$ (for FDS) can be written as

\begin{eqnarray}
E && =  \frac{V\hbar^2}{4m\pi^2}e^{-\mu}\int\limits_0^\infty {\bf k}^4
\exp\left(-\lambda\frac {\hbar^2{\bf k}^2}{2m}\right) d{\bf k} \nn \\&& =
\frac{3V}{2\lambda}e^{-\mu} \left(\frac{m}{2\pi\lambda\hbar^2}\right)^{3/2}.\label{eq:16}
\end{eqnarray}

Quantitative comparison~\cite{griffiths6} between
Eq.~(\ref{eq:16}) and with the energy of a 3D ideal gas, $E$ $=$
$3nk_BT/2$, after substituting Eq.~(\ref{eq:14}) into
Eq.~(\ref{eq:16}) will enable one to conclude $\lambda_{FDS}$ $=$
1/$k_BT$. $k_B$ is the Boltzmann constant. Applying the identical
procedure to iFDS, i.e. by employing Eqs.~(\ref{eq:10})
and~(\ref{eq:11}), then Eqs.~(\ref{eq:12}) and~(\ref{eq:13}) are
respectively rewritten as

\begin{eqnarray}
n & = &\frac {V}{2\pi^2}e^{-\mu - \lambda E_I} \int\limits_0^\infty {\bf k}^2 \exp\left(-\lambda
\frac{\hbar^2{\bf k}^2}{2m}\right) d{\bf k},\label{eq:17}
\end{eqnarray}

\begin{eqnarray}
p & = & \frac {V}{2\pi^2}e^{\mu - \lambda E_I} \int\limits_{-\infty}^0 {\bf k}^2 \exp\left(\lambda
\frac {\hbar^2{\bf k}^2}{2m}\right) d{\bf k}.\label{eq:18}
\end{eqnarray}

The respective solutions of Eqs.~(\ref{eq:17}) and~(\ref{eq:18}) are

\begin{eqnarray}
\mu + \lambda E_I & = & -\ln\left[\frac{n}{V}\left(\frac{2\pi\lambda\hbar^2}{m}\right)^{3/2}
\right],\label{eq:19}
\end{eqnarray}

\begin{eqnarray}
\mu -\lambda E_I & = & \ln\left[\frac{p}{V}\left(\frac{2\pi\lambda\hbar^2}{m}\right)^{3/2}
\right].\label{eq:20}
\end{eqnarray}

Note that Eqs.~(\ref{eq:19}) and~(\ref{eq:20}) simply imply that $\mu_e(iFDS)$ $=$
$\mu_e$ + $\lambda E_I$ and $\mu_h(iFDS)$ $=$ $\mu_h$ $-$ $\lambda E_I$. Furthermore,
using Eq.~(\ref{eq:5}), one can rewrite Eq.~(\ref{eq:16}) as

\begin{eqnarray}
E && = \frac {V\hbar^2}{4m\pi^2}  e^{-\mu -\lambda E_I} \int\limits_0^\infty {\bf k}^4
\exp\left(-\lambda\frac{\hbar^2{\bf k}^2}{2m}\right)d{\bf k} \nn \\&& =
\frac{3V}{2\lambda}e^{-\mu -\lambda E_I} \left(\frac{m}{2\pi\lambda\hbar^2}\right)^{3/2}.
\label{eq:21}
\end{eqnarray}

Again, quantitative comparison between Eq.~(\ref{eq:21}) and with
the energy of a 3D ideal gas, $E$ $=$ $3nk_BT/2$, after
substituting Eq.~(\ref{eq:19}) into Eq.~(\ref{eq:21}) will enable
one to determine $\lambda$. It is found that $\lambda$ remains the
same as 1/$k_BT$. I.e., $\lambda_{FDS}$ $=$ $\lambda_{iFDS}$ as
required by the standard FDS. Hence, the relationship between FDS
and iFDS in term of Lagrange multipliers has been derived and
shown clearly.

\subsection{2.3. Resistivity models}\lb{s-eqs}

Denoting $\mu$ $=$ $-E_F$ (Fermi level), $\lambda$ $=$ 1/$k_BT$, $\hbar^2{\bf k}^2/2m$ $=$ $E$
and substituting these into Eqs.~(\ref{eq:10}) and~(\ref{eq:11})
will lead one to write

\begin{eqnarray}
f_e(E) &=& \exp\left[\frac{E_F - E_I - E}{k_BT}\right], \label{eq:22}
\end{eqnarray}

\begin{eqnarray}
f_h(E) &=& \exp\left[\frac{E - E_I - E_F}{k_BT}\right]. \label{eq:23}
\end{eqnarray}

At this point, one might again wonder the reason for $E_I$'s
inclusion. The unique reason is that it directly determines the
kinetic energies of electrons which carry the identity of its
origin atom. Detailed experimental implications are given in the
discussion. These iFDS probability functions for electrons and
holes are unique in a sense that it allow the prediction of charge
carriers' concentrations at various $T$ and doping. It is worth
noting that, $-E_I$ in Eq.~(\ref{eq:23}) for $h$ follows naturally
from the Dirac's theory of antiparticle
interpretations~\cite{sakurai7}. Besides, the charge carriers are
not entirely free since there exist a gap-like parameter that can
be related to electrons-ion or Coulomb attraction. In fact,
application of Eqs.~(\ref{eq:22}) and~(\ref{eq:23}) in
semiconducting manganites will be explained in section IIIB. The
general equations to compute charge carriers' concentrations are
stated below,

\begin{eqnarray}
n &=& \int\limits^{\infty}_0{f_e(E)N_e(E)dE},
\label{eq:24}
\end{eqnarray}

\begin{eqnarray}
p &=& \int\limits_{-\infty}^0{f_h(E)N_h(E)dE}.
\label{eq:25}
\end{eqnarray}

Existence of $E_g$, which is the energy gap due to energy band
splitting or lattice based gap is not inserted explicitly thus it
is (if any) can be coupled with $E_I$, which is tied to ions via
Coulomb attraction. Having said that, now it is possible to obtain
the geometric-mean concentrations of electrons and holes for 1D, 2D and 3D
respectively in the forms of (assuming $n$ $\approx$ $p$)

\begin{eqnarray}
\sqrt{np}(1D) =&&
\frac{(m^*_em^*_h)^{1/4}}{\hbar}\bigg(\frac{k_BT}{2\pi}\bigg)^{1/2}  \nonumber \\&&
\times \exp\left[\frac{-E_I}{k_BT}\right],
\label{eq:26}
\end{eqnarray}

\begin{eqnarray}
\sqrt{np}(2D) & = &
\frac{k_BT}{\pi\hbar^2}\big(m_e^*m_h^*\big)^{1/2}\exp\left[\frac{-E_I}{k_BT}\right],
\label{eq:27}
\end{eqnarray}

\begin{eqnarray}
\sqrt{np}(3D) =&&
2\left[\frac{k_BT}{2\pi\hbar^2}\right]^{3/2}(m^*_em^*_h)^{3/4}  \nn \\&&
\times \exp\left[\frac{-
E_I}{k_BT}\right]. \label{eq:28}
\end{eqnarray}

In which

\begin{eqnarray}
n(3D) =&&
2\left[\frac{k_BT}{2\pi\hbar^2}\right]^{3/2}(m^*_e)^{3/2}  \nn \\&&
\times \exp\left[\frac{E_F -
E_I}{k_BT}\right]. \label{eq:29}
\end{eqnarray}

\begin{eqnarray}
p(3D) =&&
2\left[\frac{k_BT}{2\pi\hbar^2}\right]^{3/2}(m^*_h)^{3/2}  \nn \\&&
\times \exp\left[\frac{-E_F -
E_I}{k_BT}\right]. \label{eq:30}
\end{eqnarray}

It is apparent that $n$(1D,2D,3D) $\propto$ $\exp [(E_F -
E_I)/k_BT]$ and $p$(1D,2D,3D) $\propto$ $\exp [(-E_F - E_I)/k_BT]$
while the rest are just typical physical constants except $T$. As
such, the constant, $E_F$ in the exponential term will appear
where appropriate if one assumes $n$ $\gg$ $p$ or $p$ $\gg$ $n$
instead of $\sqrt{np}$. The DOS, $N(E,1D)$ $=$
$(E^{-1/2}\sqrt{m^*/2})/\pi\hbar$, $N(E,2D)$ $=$ $m^*/\pi\hbar^2$
and $N(E,3D)$ $=$ $(E^{1/2}/2\pi^2)(2m^*/\hbar^2)^{3/2}$ were
employed in which, $m^*$ is the effective mass. Consequently, the
resistivity models for 1D, 2D and 3D can be derived from $\rho$
$=$ $m/ne^2\tau$ by taking 1/$\tau$ $=$ $A_{D=1,2,3}T^2$ as a
consequence of electron-electron scattering and $m = m^*_e \approx
m^*_h \approx (m^*_em^*_h)^{1/2}$. The subscript $D = 1, 2, 3$
represents dimensionality, i.e. $A_1$, $A_2$ and $A_3$ are
$T$-independent scattering rate constants in 1D, 2D and 3D
respectively. $\tau$ denotes scattering rate due to
electron-electron scattering in the absence of {\bf H}. Therefore,
the respective $\rho(T,E_I)$ are given by

\begin{eqnarray}
\rho(1D)  =&&
\frac{A_1\hbar(m^*_em^*_h)^{1/4}}{e^2}\bigg(\frac{2\pi}{k_B}\bigg)^{1/2}  \nonumber \\&&
\times T^{3/2}\exp\left[\frac{E_I}{k_BT}\right],
\label{eq:31}
\end{eqnarray}

\begin{eqnarray}
\rho(2D) & = &
\frac{A_2\pi\hbar^2}{e^2k_B}T \exp\left[\frac{E_I}{k_BT}\right],
\label{eq:32}
\end{eqnarray}

\begin{eqnarray}
\rho(3D) & = &
\frac{A_3}{2e^2}\left[\frac{2\pi\hbar^2}{k_B}\right]^{3/2}(m^*_e
m^*_h)^{-1/4}\nonumber \\&&
\times T^{1/2} \exp\left[\frac{E_I}{k_BT}\right].
\label{eq:33}
\end{eqnarray}

In which

\begin{eqnarray}
\rho_e(3D) & = &
\frac{A_3}{2e^2}\left[\frac{2\pi\hbar^2}{k_B}\right]^{3/2}(m^*_e)^{-1/2}\nonumber \\&&
\times T^{1/2} \exp\left[\frac{E_I - E_F}{k_BT}\right].
\label{eq:34}
\end{eqnarray}

\begin{eqnarray}
\rho_h(3D) & = &
\frac{A_3}{2e^2}\left[\frac{2\pi\hbar^2}{k_B}\right]^{3/2}(m^*_h)^{-1/2}\nonumber \\&&
\times T^{1/2} \exp\left[\frac{E_I + E_F}{k_BT}\right].
\label{eq:35}
\end{eqnarray}

\subsection{2.4. Hall resistance}\lb{s-eqs}

The equations of motion for charge carriers under the influence of
static {\bf H} and electric field ({\bf E}) can be written in an identical
fashion as given in Ref.~\cite{kittel8}, which are given by

\begin{eqnarray}
m\left[ \frac{d}{dt} + \frac{1}{\tau_H} \right]v_y = -e{\bf E}_y - e{\bf H}_xv_z,\label{eq:36}
\end{eqnarray}

\begin{eqnarray}
m\left[ \frac{d}{dt} + \frac{1}{\tau_H} \right]v_z = -e{\bf E}_z + e{\bf H}_xv_y.\label{eq:37}
\end{eqnarray}

The subscripts $x$, $y$ and $z$ represent the axes in $x$, $y$ and
$z$ directions while the scattering rate, 1/$\tau_H$ $=$
$A_{D=2,3}^{(H)}T^2$ in which $A_{D=2,3}^{(H)}$ may not be
necessarily equals to $A_{D=2,3}$, though both $A_{D=2,3}$ and
$A_{D=2,3}^{(H)}$ are independent of $T$. $A^{(H)}$ and $\tau_H$ denote the
$T$-independent scattering rate constant and scattering rate respectively
with applied {\bf H}. In a steady state of a static {\bf
H} and {\bf E}, $dv_z/dt$ = $dv_y/dt$ = 0 and $v_z$ = 0 hence,
{\bf E}$_z$ can be obtained from

\begin{eqnarray}
{\bf E}_z = -\frac{e{\bf H}_x{\bf E}_y\tau_H}{m}.\label{eq:38}
\end{eqnarray}

In addition, it is further assumed
that $\rho_x(T)$ $=$ $\rho_y(T)$ $=$ $\rho_{z}(T)$ $=$ $\rho(T)$. $R_H^{(z)}$ is defined as
{\bf E}$_z$/$j_y${\bf H}$_x$, $j_y$ $=$ {\bf E}$_y$/$\rho(T)$ and $\tan\theta_H^{(z)}$ $=$
{\bf E}$_z$/{\bf E}$_y$. Parallel to this,

\begin{eqnarray}
R_H^{(z)} = \frac{\tan \theta_H^{(z)}\rho(T)}{{\bf H}_x}.\label{eq:39}
\end{eqnarray}

$j_y$ is the current due to charge carriers' motion along $y$-axis
and $\theta_H^{(z)}$ is the Hall angle. Furthermore, $\tan
\theta_H^{(z)}$ can be rewritten as $-e{\bf H}_x\tau_H/m$.
Therefore, it is easy to surmise that $\cot\theta_H^{(z)}$
$\propto$ $T^2$. After employing Eq.~(\ref{eq:32}) with $E_F$ ($n$
$\gg$ $p$) and Eq.~(\ref{eq:34}), then one can respectively arrive
at

\begin{eqnarray}
R_H^{(2D)} &=& -\frac{A_2\pi \hbar^2}{A_2^{(H)}em^*_ek_B}T^{-1}
\exp \left[ \frac{E_I - E_F}{k_BT} \right],\label{eq:40}
\end{eqnarray}

\begin{eqnarray}
R_H^{(3D)} &=& -\frac{A_3}{2A_3^{(H)}e^2(m^*_e)^{3/2}} \left[\frac{2\pi\hbar^2}{k_B} \right]
^{-3/2}\nonumber \\&&
\times T^{-3/2}\exp \left[ \frac{E_I - E_F}{k_BT} \right].\label{eq:41}
\end{eqnarray}

The negative charges in Eqs.~(\ref{eq:38}),~(\ref{eq:40}) and~(\ref{eq:41}) are due to
the assumption that the charge carriers are electrons.
Note that $R_H$($T$,1D) for any given samples that exhibit purely 1D conduction is
obviously irrelevant or simply, could not be derived with above procedures, since Hall effect
requires at least 2D conduction.

\subsection{2.5. Special cases}\lb{s-eqs}

There are nowhere in the above derivations that take into account
any free electrons and $T$-dependence of electron-phonon
scattering. Hence, the models derived thus far are obviously not
suitable for such applications except for semiconducting free
electrons above conduction band. In this case, Eqs.~(\ref{eq:24})
and~(\ref{eq:25}) should be integrated from $E_g$ $\to$ $\infty$
and 0 $\to$ $-\infty$ respectively after replacing $E_I$ $=$ 0 in
Eqs.~(\ref{eq:22}) and~(\ref{eq:23}). As for metals with free
electrons and strong phonon contributions, it is advisable to
switch to the well known Bloch-Gr\"{u}neisen formula given
by~\cite{tu9},

\begin{eqnarray}
\rho(T,3D) &=& \rho_0 + \lambda_{tr} \frac{128 \pi m^*
(k_BT)^5}{ne^2(k_B\Theta_D)^4} \nn \\&& \times
\int\limits_0^{\Theta_D/2T} \frac {x^5}{\sinh^2x} dx.\label{eq:42}
\end{eqnarray}

$\lambda_{tr}$ $=$ electron-phonon coupling constant, $\rho_0$ $=$
$\rho(T = 0)$, $m^*$ = average effective mass of the occupied
carrier states, $\Theta_D$ $=$ Debye temperature, $n$ $=$ free
electrons concentrations. As a matter of fact, one should {\it
not} be encouraged to substitute any of the
Eqs.~(\ref{eq:26})$-$(\ref{eq:30}) for $n$ into Eq.~(\ref{eq:42})
just to capture the electron-phonon scattering because the
scattering of free electrons considered in Eq.~(\ref{eq:42}) may
not be compatible with {\it gapped}-electrons' scattering of iFDS,
unless proven otherwise.

\subsection{2.6. The rationale behind iFDS}\lb{s-eqs}

Recall that $E_I$ is the gap from electron-ion attraction along
the easiest path while $E_g$ is the averaged gap from the energy
band splitting. $E_g$ from the theory of electronic band structure
(TEBS) does not give the correct prediction on how certain ionic
substitutions affect the transport properties of strongly
correlated systems. As I have pointed out in the last paragraph of
the introduction, iFDS follows a brand new approach instead of the
common TEBS, which is not suitable and/or reliable for the
strongly correlated systems namely, superconductors, ferromagnets
and ferroelectrics. In fact, application of iFDS in the normal
state of high-$T_c$ superconductors have been carried out while
its consequences in ferroelectrics has been reported as
well~\cite{andrew3}. Usually, one needs to incorporate TEBS into
FDS to take the lattice properties into account to expose the
transport mechanism and one can also include the defects so as to
arrive at a comprehensive theory. However, the magnitude of
lattice effects and anisotropies calculated from TEBS is obviously
not suitable for the transport theory because the charge carriers
move along the easiest path complying the principle of least
action. This is why TEBS-FDS is not reliable when it comes to the
transport properties of strongly correlated systems. It is also
important to realize that for simple systems, the lattice
properties are isotropic therefore the magnitude of say, $E_g$
calculated from TEBS is comparable with $E_g$ from the easiest
path, thus simple systems are very well described by TEBS-FDS.
Unlike TEBS-FDS, iFDS takes care of the crystalline lattice via
$E_I$ in accordance with the easiest path (transport measurements)
since any changes to lattice's crystallinity due to substitution
will be mirrored through the easiest path's $E_I$ and subsequently
shall be picked up by the electrons, influencing its transport
mechanism. Whereas, the TEBS-FDS theory assumes that all the
lattice effects including $E_g$ and its magnitudes are isotropic,
which will be imposed on the electrons unnecessarily. In other
words, it is logically desirable to let the electrons reveal the
obstacles via iFDS rather than forcing $E_g$ onto the electrons in
order to calculate the electronic transport properties.

I would like to stress here that if one substitutes $E_I$ with
$E_g$ in Eq.~(\ref{eq:33}) to comply with the standard FDS, then
one has to feed the value of $E_g$ from elsewhere (such as TEBS)
in order to predict the transport properties and apparently, TEBS
is not suitable for the reasons described above. Hence, the
Eqs.~(\ref{eq:26})$-$(\ref{eq:35}) and
Eqs.~(\ref{eq:40})$-$(\ref{eq:41}) can be directly employed to
analyze the doping effect on the transport properties of
ferromagnets above $T_C$, be it diluted (DMS) or concentrated
(manganites).  Take note that the proposed theory based on iFDS is
in its infancy, thus the correct method to insert defects via
$E_I$ will be developed later on. Consequently, the application of
iFDS is to be confined to pure crystalline and polycrystalline
samples. The above-stated arguments do not invalidate the
importance of TEBS since the lattice properties and its
consequences on electronic structure are indeed invaluable to
derive lattice based Hamiltonian.

\section{3. Discussion}\lb{s-eqs}

\subsection{3.1. Simulated curves}\lb{s-eqs}

Figures~\ref{fig1}a, 1b and 1c illustrate the variation of
$\rho(T)$ from Eqs.~(\ref{eq:31})$-$(\ref{eq:33}) respectively
with conduction dimensionality and doping parameter ($E_I$). The
proportionality of Eqs.~(\ref{eq:31})$-$(\ref{eq:33}) with $E_I$
is similar to Eqs.~(\ref{eq:34}) and~(\ref{eq:35}), thus it does
not matter whether one chooses $\sqrt{np}$, $n$ or $p$ in the
subsequent analysis as well as for manganites. One can also
identify the $\rho(T)$ transition from metallic $\to$
semiconducting  conduction with increasing $E_I$. It is also worth
noting that $\rho(T$,1D), $\rho(T$,2D) and $\rho(T$,3D) are
$\propto$ $T^{3/2}$, $T$ and $\sqrt{T}$ respectively if and only
if $E_I$ $\ll$ $T$. Another point worth to extract from these
curves are the relationship between $T_{crossover}$ and $E_I$,
where $T_{crossover}$ $<$ $E_I$ for 1D, $T_{crossover}$ $=$ $E_I$
for 2D and $T_{crossover}$ $>$ $E_I$ for 3D. Apparently, these
relations are again due to the proportionalities of $\rho(T$,1D)
$\propto$ $T^{3/2}$, $\rho(T$,2D) $\propto$ $T$ and $\rho(T$,3D)
$\propto$ $\sqrt{T}$. Figures~\ref{fig2}a and 2b plot the
simulated $R_H(T)$ curves in 2D and 3D as well as at different
$E_I$(0 K, 150 K, 310 K) that follow from Eqs.~(\ref{eq:40})
and~(\ref{eq:41}). There are no significant differences of $R_H$
between 2D and 3D since $R_H$(2D) and $R_H$(3D) are $\propto$ to
$T^{-1}$ and $T^{-3/2}$ respectively. Besides, the $T$ from
$\exp{[E_I/T}]$ also inversely proportional to both $R_H$(2D) and
$R_H$(3D). These scenarios will always lead $R_H$ to increase with
lowering $T$ without any observable $T_{crossover}$ regardless of
$E_I$'s magnitude, unlike $\rho(T)$. It is convenient to directly
quantify $\rho(T)$ variation with doping by relating $E_I$ as a
doping parameter, as will be discussed in the following paragraph
for manganites at $T$ $>$ $T_C$.

\subsection{3.2. Manganites}\lb{s-eqs}

As stated previously, Jonker and van Santen~\cite{jonker7,santen8}
suggested that FM is due to indirect coupling of $d$-shells via
conducting electrons. Subsequently, Zener~\cite{zener5,zener6},
Anderson and Hasegawa~\cite{anderson14} have provided sufficient
theoretical backgrounds on DEM. However, this paper will not
discuss the property of DEM below $T_C$, instead the electrical
properties above $T_C$ (paramagnetic phase) will be addressed in
detail in which, DEM is employed at $T$ $<$ $T_C$ (ferromagnetic
phase). It is interesting to observe reduced $\rho(T)$ and
increased Curie temperature ($T_C$) at higher ${\bf H}$ for
La$_{1-x}$Ca$_x$-Sr$_x$MnO$_3$ compounds
~\cite{mukovskii15,mahendiran16,gupta17}. The results that larger
{\bf H} giving rise to conductivity at $T$ $\geq$ $T_C$ point
towards the enhancement of conductivity from DEM where the
exponential increase of $\rho(T)$ is suppressed with larger {\bf
H}. Additionally, small polarons have also been
attributed~\cite{billinge18} to play a significant role on
$\rho(T)$ at $T$ $>$ $T_C$. As a matter of fact, this polaronic
effect is naturally captured by Eqs.~(\ref{eq:33})$-$(\ref{eq:35})
in which the gap-parameter, $E_I$ that represents electron-ion
attraction is also a parameter that measures the combination of
electrons and its strain field due to neighboring ions, which is
nothing but polarons. The absolute value of $E_I$ can be obtained
from~\cite{kittel8}

\begin{eqnarray}
E_I &=& \frac{e^2}{8\pi\epsilon\epsilon_0r_B}.\label{eq:43}
\end{eqnarray}

$\epsilon$ and $\epsilon_0$ are the dielectric constant and
permittivity of free space respectively, $r_B$ is the Bohr radius.
Furthermore, the decrement of $E_I$ with {\bf H} indicates that
$r_B$ increases with {\bf H}. Identical relationship was also
given between polaronic radius, $r_p$ and $E_p$ by Banerjee {\it
et al}.~\cite{banerjee19}. Actually, Millis {\it et
al}.~\cite{millis10} have somewhat proved the inadequacy of DEM
alone to describe $\rho(T,doping$,{\bf H}) and reinforced the need
to include small polarons as a consequence of Jahn-Teller (JT)
splitting of Mn$^{3+}$ ions. This statement was further justified
by experimental work of Banerjee {\it et
al}.~\cite{banerjee19,banerjee21,banerjee22} in which, they have
established the existence of small polarons at $T$ $>$ $T_C$ in
La$_{0.5}$Pb$_{0.5}$Mn$_{1-x}$Cr$_x$O$_3$ for $x$ $=$ 0 $\to$ 0.45
using thermoelectric power and positron annihilation lifetime
measurements. Banerjee {\it et al}. suggested that the
substitution of Cr$^{3+}$ into Mn sites localizes $e^1_g$
electrons that gives rise to $\rho(T)$
~\cite{banerjee19,banerjee21,banerjee23}. However, the increment
of activation energy, $E_p$ with $x$ is still unclear in term of
$e^1_g$ electrons' localization. On the other hand, $\rho(T)$
(Eqs.~(\ref{eq:33})$-$(\ref{eq:35})) based on iFDS could explain
the increment of $E_I$ with $x$ as well as the structural changes
accompanied by this doping, which is due to the fact that valence
state of Cr and Mn may change with doping that can be calculated
with Eq.~(\ref{eq:44}) as will be shown later.

In addition, Moskvin~\cite{moskvin24,moskvin25} reinforces the
importance of considering different charge distribution in
MnO$_6$, Mn and Oxygen instead of just considering the DEM.
Parallel to this, doping-friendly resistivity models derived from
iFDS will indeed lead to identical conclusions with Billinge {\it
et al}. and Moskvin since iFDS naturally considers polarons and
the substitution of Ca$^{2+}$ or Cr$^{2+,3+,4+}$ into
La$_{1-x}$Ca$_x$Mn$_{1-y}$Cr$_y$O$_3$ system does indeed modify
the overall charge distribution in accordance with the valence
state of Ca, La, Mn and Cr. Add to that, Louca and
Egami~\cite{louca26} invoked the JT distortion to describe the
effect of lattice on $T_C$. They have utilized the results of
pulsed neutron-diffraction experiments to conclude that the
variation in Mn-O bond length with Sr substitution in
La$_{1-x}$Sr$_x$MnO$_3$ compound can be related to JT. Again, note
here that the change of Mn-O length with Sr substitution implies
the valence state of Mn varies with doping. In an identical
compound of La$_{1-x}$Ca$_x$MnO$_3$, substitution of Ca into La,
will have to satisfy the inequality of average $E_I$ between
Ca$^{2+}$ (867 kJmol$^{-1}$) and La$^{3+}$ (1152 kJmol$^{-1}$)
i.e., $E_I$(La$^{3+}$) $>$ $E_I$(Ca$^{2+}$). Here, one can easily
fix the valence state of Ca$^{2+}$ and La$^{3+}$ as noted. As a
consequence, this will ease the prediction of $\rho(T)$ with
doping. If one of the ions is multivalence, then the linear
algebraic equation as given below must be used to predict the
valence state of the multivalence ion from $\rho(T)$
curves~\cite{arulsamy2,arulsamy2a}.

\begin{eqnarray}
\frac{\delta}{j}\sum^{z+j}_{i=z+1}{E_{Ii}} +
\frac{1}{z}\sum^{z}_{i=1}{E_{Ii}} & = &
\frac{1}{q}\sum^{q}_{i=1}{E_{Ii}}. \label{eq:44}
\end{eqnarray}

The first term, $\frac{\delta}{j}\sum^{z+j}_{i=z+1}{E_{Ii}}$ above
has $i$ $=$ $z$ + 1, $z$ + 2,..., $z$ + $j$ and $j$ $=$ 1, 2,
3,.... It is solely due to multivalence ion for example, assume
Mn$^{3+,4+}$ is substituted with Nd$^{3+}$
(La$_{0.7}$Ca$_{0.3}$Mn$_{1-x}$Nd$_x$O$_3$) hence from
Eq.~(\ref{eq:44}), the first term is due to Mn$^{4+}$ ion's
contribution or caused by reaction of the form Mn$^{3+}$ $-$
electron $\to$ Mn$^{4+}$ (4940 kJmol$^{-1}$), hence $j$ is equals
to 1 in this case and $\delta$ represents the additional
contribution from Mn$^{4+}$. The second ($i$ $=$ 1, 2, 3, ...,
$z$) and last ($i$ $=$ 1, 2, 3, ..., $q$) terms respectively are
due to reaction of the form Mn $-$ 3(electrons) $\to$ Mn$^{3+}$
and Nd $-$ 3(electrons) $\to$ Nd$^{3+}$. Recall that $q$ = $z$ =
3+ and $i$ = 1, 2, 3,... represent the first, second, third, ...
ionization energies while $j$ = 1, 2, 3, ... represent the fourth,
fifth, sixth, ... ionization energies. Therefore, $z$ + $\delta$
gives the minimum valence number for Mn which can be calculated
from Eq.~(\ref{eq:44}). Now, it is possible to explain the doping
effect in La$_{0.5}$Pb$_{0.5}$Mn$_{1-x}$Cr$_x$O$_3$
system~\cite{banerjee19,banerjee21}. The inequalities of $E_I$s
are given as Mn$^{3+}$ (1825 kJmol$^{-1}$) $>$ Cr$^{3+}$ (1743
kJmol$^{-1}$) and Mn$^{4+}$ (2604 kJmol$^{-1}$) $>$ Cr$^{4+}$
(2493 kJmol$^{-1}$). These relations strongly indicate that
$\rho(T)$ should decrease with Cr$^{3+}$ content contradicting
with experimental data from Refs.~\cite{banerjee19,banerjee21}.
The only way to handle this situation is to use Eq.~(\ref{eq:44})
so as to calculate the minimum valence state of Cr$^{3 + \delta}$,
which is 3.033+. Actually, the valence state of Cr that
substitutes Mn$^{3+}$ is Cr$^{> 3.033}$ and of course, the valence
state of Mn is fixed to be 3+. There is no need to vary it because
$\rho(T)$ was found to increase with Cr
content~\cite{banerjee19,banerjee21}. Actually, the existence of
Mn$^{2+}$ is unlikely and negligible, while Mn$^{4+}$ (if any)
will further increase the value of $\delta$ in Cr$^{3 + \delta}$,
in which both scenarios comply with iFDS via Eq.~(\ref{eq:44}) and
the resistivity measurements~\cite{banerjee19,banerjee21}.

The effect of hydrostatic (external) pressure ($P$) and chemical
doping (internal $P$) on metal-insulator transition of
Pr$_{0.7}$Ca$_{0.2}$Sr$_{0.1}$MnO$_3$,
Pr$_{0.7}$Ca$_{0.21}$Sr$_{0.09}$MnO$_3$,
Pr$_{0.58}$La$_{0.12}$Ca$_{0.3}$MnO$_3$ and
Pr$_{0.54}$La$_{0.16}$Ca$_{0.3}$MnO$_3$ systems were reported by
Medvedeva {\it et al}.~\cite{med27}. It is found that $\rho(T)$
and $T_C$ are observed to be decreased and increased respectively
with increasing $P$ ranging from 0 $\to$ 15 kbar. As anticipated,
$\rho(T)$ above and below $T_C$ have been decreased significantly
with $P$ i.e. $P$ affects both Mn-O-Mn bond angle and length.
Hence, it is apparent that doping and $P$ give rise to the
variation in the valence state of Pr and Mn in order to achieve a
certain crystal structure and simultaneously increase the number
of charge carriers. As such, changes in $\rho(T)$ above $T_C$ can
be very well accounted for with
Eqs.~(\ref{eq:28})$-$(\ref{eq:30}),~(\ref{eq:33})$-$(\ref{eq:35})
where $P$ reduces $E_I$ (increases $r_B$) of certain ions in a
similar fashion to doping (internal $P$). However, this paper as
stated earlier does not attempt to describe the correct
mechanism(s) involved below $T_C$ and its variations with $P$ and
doping. In short, at $T$ $<$ $T_C$, the interactions among DEM
with JT or polarons or all may play a significant role as
suggested by Billinge {\it et al}.~\cite{billinge18} and Medvedeva
{\it et al}.~\cite{med27}.

\subsection{3.3. Diluted magnetic semiconductors}\lb{s-eqs}

It is quite straight forward to extend iFLT described above to
DMS. Firstly, Oiwa {\it et al}.~\cite{oiwa}, Matsukura {\it et
al}.~\cite{matsu}, Dietl {\it et al}.~\cite{dietllast} and Iye
{\it et al}.~\cite{iye} have done detailed investigations on
Ga-As-Mn based DMS. The variation of electrical resistivities with
temperature and doping for Ga$_{1-x}$Mn$_x$As in
Refs~\cite{oiwa,matsu,dietllast} (sample A1) differ slightly from
Ref~\cite{iye} (sample B2). It has been reported that the maximum
$T_C$ is $\sim$ 70 K for Ga$_{0.957}$Mn$_{0.043}$As~\cite{iye}
whereas Ga$_{0.947}$Mn$_{0.053}$As had a $T_C$ of $\sim$ 100 K.
Apart from $T_C$s, there is a switch-over in the variation of
$\rho(T,x)$ where $\rho(T,x)$ decreases with $x$ initially before
switching to increasing $\rho(T,x)$ with $x$. $\rho(T,x)$'s
variation for sample A1 is $\rho(T,0.015)$ $>$ $\rho(T,0.022)$ $>$
$\rho(T,0.071)$ $>$ $\rho(T,0.053)$ $>$ $\rho(T,0.043)$ $>$ {\bf
$\rho(T,0.035)$} while for sample B2, $\rho(T,0.015)$ $>$
$\rho(T,0.022)$ $>$ $\rho(T,0.071)$ $>$ $\rho(T,0.035)$ $>$
$\rho(T,0.043)$ $>$ {\bf $\rho(T,0.053)$}. The initial reduction
of $\rho(T)$ with $x$ is due to $E_I$ (Mn$^{3+}$ = 1825
kJmol$^{-1}$) $<$ $E_I$ (Ga$^{3+}$ = 1840 kJmol$^{-1}$), i.e. it
supports that Mn$^{3+}$ substitutes at Ga$^{3+}$ sites. However,
$\rho(T)$'s switch-over (from decreasing to increasing with $x$)
at critical concentrations, $x_{c1}$ = 0.035 and $x_{c2}$ = 0.053
for samples A1 and B2 respectively can only be explained if one
employs the mechanism proposed by Van Esch {\it et
al.}~\cite{esch13,esch14} and Ando {\it et al.}~\cite{ando22} in
which, Mn$^{6As}$ formations is substantial above $x_c$ in such a
way that Mn$^{3+}$ ions do not substitute Ga$^{3+}$ ions.
Therefore, $\rho(T)$ will be influenced by Mn$^{6As}$ clusters,
defects and Ga-Mn-As phase simultaneously significantly above
$x_c$. This is indeed in fact in accordance with iFDS based
resistivity models since if one assumes Mn$^{2+}$ substitutes
Ga$^{3+}$, then $\rho(T)$ should further decrease with $x$ without
any switch-overs. Hence, the formation of Mn$^{6As}$ with $x$ is
inevitable in III-V Ga$_{1-x}$Mn$_x$As DMS. Again, $T_C$ of DMS is
also directly proportional to \textbf{H}, identical to manganites.
All the values of $E_I$ discussed above (including in section
IIIB) were averaged in accordance with $E_I [X^{z+}] =
\sum_{i=1}^z\frac{E_{Ii}}{z}$ and should not be taken literally
since those $E_I$s are not absolute values. The absolute values
need to be obtained from the $r_B$ dependent $E_I$ equation stated
earlier~\cite{andrew,andrew4}. Prior to averaging, the 1$^{st}$,
2$^{nd}$, 3$^{rd}$ and 4$^{th}$ ionization energies for all the
elements mentioned above were taken from Ref.~\cite{web28} and the
predictions stated above are only valid for reasonably pure
materials without any significant impurity phases as well as with
minimal grain boundary effects.

\section{4. Conclusions}\lb{s-eqs}

In conclusion, the ionization energy based Fermi-Dirac statistics
is useful to estimate the transitional progress of
$\rho(T,doping,pressure$,{\bf H}) from metallic $\to$
semiconducting or vice versa in both manganites and diluted
magnetic semiconductors. This is made possible by an additional
unique constraint, which is nothing but the ionization energy that
captures the electron's kinetic energy and maps it to its origin
atom. The relation of Lagrange multipliers ($\lambda$ and $\mu$)
between FDS and iFDS have been derived explicitly solely to
flush-out any misinterpretations that will lead to further
complications in describing experimental data on semiconducting
ferromagnets and DMS at $T$ $>$ $T_C$. The presented iFDS model
however, does not admit completely free-electrons and strong
electron-phonon scattering. Importantly, $E_I$ captures the
polaronic effect quite naturally to explain the paramagnetic phase
electrical properties of the said materials accurately.

\section*{Acknowledgments}

The author is grateful to Arulsamy Innasimuthu, Sebastiammal
Innasimuthu, Arokia Das Anthony and Cecily Arokiam of CMG-A for
the financial aid. ADA also thanks Prof. Feng Yuan Ping for his
support.

\begin{figure}
\caption{Simulated $\rho(T,E_I)$ curves from
Eqs.~(\ref{eq:31}),~(\ref{eq:32}) and~(\ref{eq:33}) are depicted
in a), b) and c) respectively. These curves capture 1D, 2D and 3D
conduction respectively at various $E_I$ between 0 K and 310 K.
The intrinsic $T$-dependence among 1D, 2D and 3D conduction are
$T^{3/2}$, $T$ and $\sqrt{T}$ respectively for $T$ $>$
$T_{crossover}$ whereas all $\rho(T,E_I)$ behave identically as
$\rho$ $\propto$ $\exp (1/T)$ at $T$ $<$ $T_{crossover}$ since
$T_{crossover}$ is less than the numerator in the exponential
terms. The latter proportionalities can be verified from
Eqs.~(\ref{eq:31}),~(\ref{eq:32}) and~(\ref{eq:33}).} \label{fig1}
\end{figure}

\begin{figure}
\caption{Simulated 2D and 3D curves of Hall resistances
($R_H(T,E_I$) from Eqs.~(\ref{eq:40}) and~(\ref{eq:41}) are given
in a) and b) respectively. All curves for both dimensionalities
(2D and 3D) are also plotted at $E_I$ = 0 K, $E_I$ = 150 K and
$E_I$ = 310 K. Note that $R_H(T,E_I)$ curves are inversely
proportional to $T$ regardless of the magnitude of $T_{crossover}$
that can be easily verified from Eqs.~(\ref{eq:40})
and~(\ref{eq:41}). Therefore, one could not excerpt sufficient
quantitative differences between 2D and 3D $R_H(T,E_I)$ as opposed
to $\rho(T,E_I)$.} \label{fig2}
\end{figure}

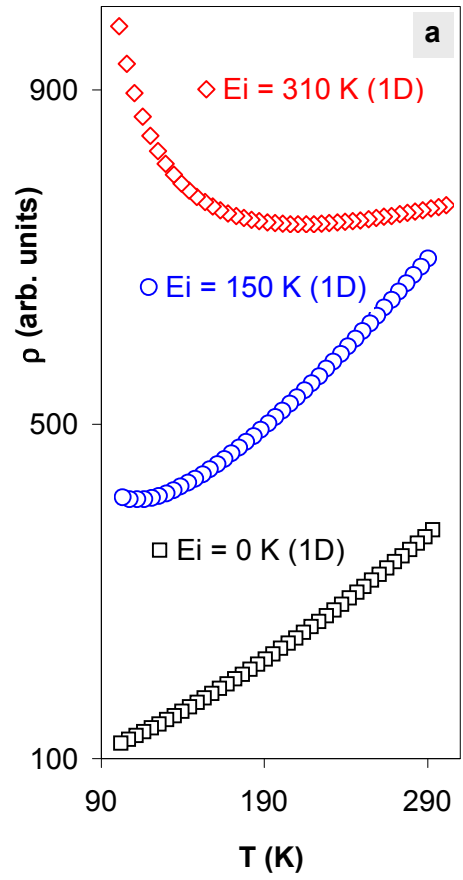
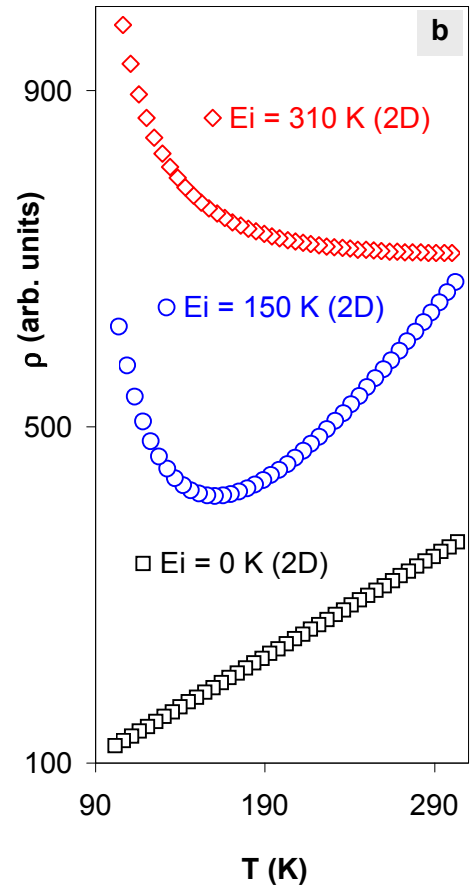
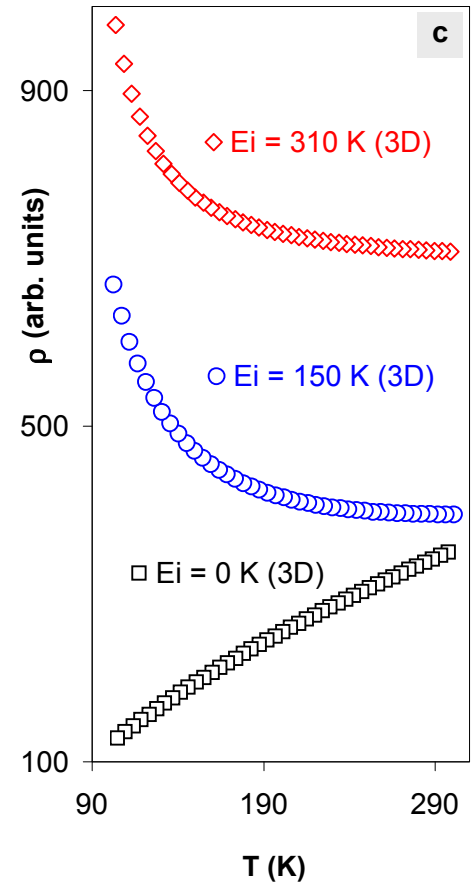
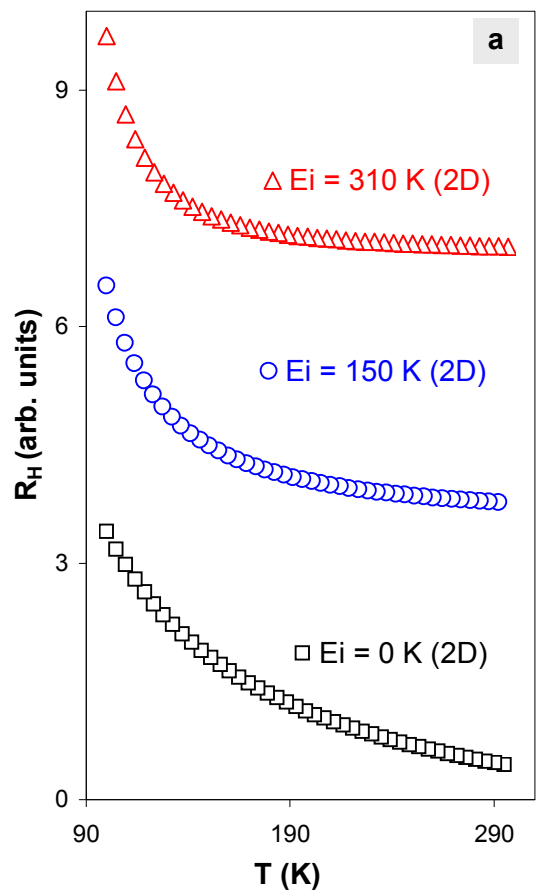
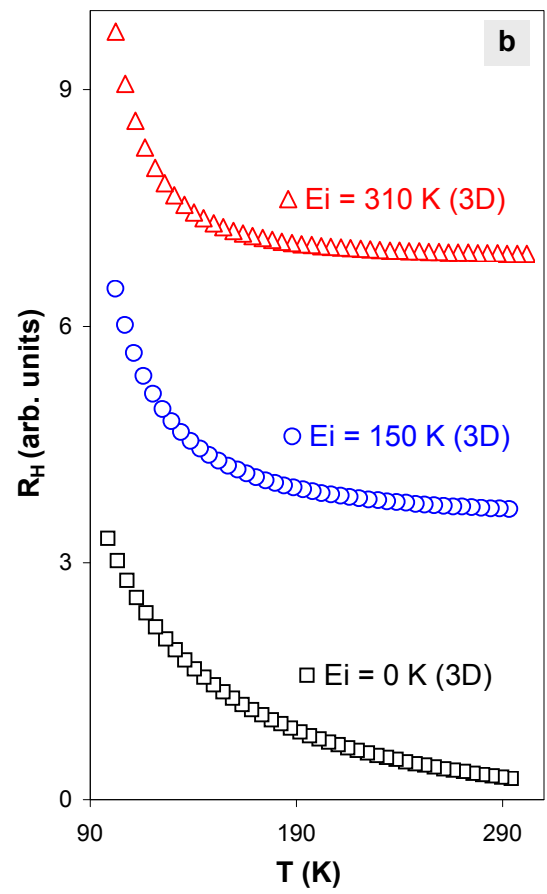
\end{document}